# Strain-Driven Electronic and Catalytic Modulation of g-$C_3N_4$/GeS van der Waals heterostructure for Photocatalytic Water Splitting

Soumendra Kumar Das[1], Smruti Ranjan Parida[1], Tapas Kumbhakar[1], Prasanjit Samal[2] and Sridhar Sahu[1*]

*[1] Department of Physics, Indian Institute of Technology (Indian School of Mines) Dhanbad, Dhanbad-826004, Jharkhand, India*

*[2]School of Physical Sciences, National Institute of Science Education and Research (NISER) Bhubaneswar, HBNI, Jatni, Khurda-752050, Odisha, India*

**corresponding authors: e-mail: sridharsahu@iitism.ac.in*

(Dated: August 29, 2026)

**Abstract**

In the present study, we report the construction of a two-dimensional g-$C_3N_4$/GeS heterostructure and a comprehensive analysis to evaluate its photocatalytic performance for water splitting, HER and OER under the influence of mechanical strain. First-principles calculations predict that the complementary band edge positions of the GeS and g-$C_3N_4$ monolayers form a type-II band alignment, which improves the photocatalytic efficiency by reducing the recombination rate and enhancing the separation between photogenerated electrons and holes. The heterostructure exhibits a reduced band gap of 2.17 eV, as compared to that of the free-standing monolayers of g-$C_3N_4$ (2.81 eV) and GeS (3.31 eV), respectively. Application of biaxial strain modifies the electronic structure significantly and reduces the band gap systematically to 1.6 eV at 5% tensile strain. The calculated band edges straddle the standard water redox potential value, enabling favourable water splitting to execute the hydrogen evolution and oxygen evolution reactions, which is further enhanced by the built-in electric field between the monolayers. The calculated Gibbs' free energy change for the hydrogen evolution reaction ($\Delta G_{HER}$) indicates a low value of 0.2 eV for the heterostructure, which further decreases to $\mp 0.1\ eV$ with +1% and +2% tensile strain, and increases to +0.38 eV for 3% tensile strain, respectively. Similarly, the OER overpotential for the heterostructure was around 2.17 V, which gradually decreases to 1.06 V, 1.02V, and 0.97 V with 1%, 2%, and 3% tensile strain, respectively, indicating improvement in the OER activity. The optical absorption edge falls in the visible region, indicating the possibility of using maximum visible light for water splitting. The thermodynamic stability of the monolayers and heterostructure was verified through ab initio molecular dynamics simulations, indicating minimal changes in total energy. Our calculations provide valuable insights for using g-$C_3N_4$/GeS as an efficient photocatalyst for sustainable hydrogen production and shed light on designing artificial heterostructures for future energy applications.

**Keywords:** 2D Materials, Photocatalytic water splitting, Type-II band alignment, Strain-engineering, Hydrogen evolution reaction, Oxygen evolution reaction,

## 1. INTRODUCTION

The excessive use of fossil fuels and growing environmental concerns have led to a global push for sustainable, clean energy alternatives [1, 2]. Among the many possibilities, hydrogen stands out as a highly attractive energy carrier due to its high energy density, environmental friendliness, and versatility for industrial applications[3, 4]. Among various methods, photocatalytic water splitting, which uses solar energy to produce hydrogen, is a green, potentially cost-effective approach for large-scale hydrogen generation, significantly contributing to the development of a low-carbon energy infrastructure[5]. Since the pioneering work on $TiO_2$ in 1972, there has been a concentrated effort to identify and develop efficient photocatalysts with desirable features, such as narrow band gaps, favourable band edge positions, strong redox capabilities, and efficient charge separation[6, 7]. Conventional semiconductor photocatalysts often exhibit limited visible-light absorption and high charge-carrier recombination rates, which reduce their practical efficiency for hydrogen production [8]. On the other hand, the current photocatalytic devices are too expensive and ineffective for producing hydrogen on a commercial scale. Many two-dimensional (2D) materials have been designed and predicted as possible photocatalysts for water splitting since the development of high-performance computers [9]. Large specific surface area, short charge-carrier pathways, and variable electrical properties are among the key benefits 2D nanomaterials offer for overall water splitting, thereby improving photocatalytic performance [10-12].

Although graphene, a two-dimensional carbon-based material, has interesting mechanical and electrical properties, its zero band gap prevents its use in photocatalysis [13, 14]. Other newly developed 2D materials, however, have superior photocatalytic qualities[11, 15]. For example, $C_3N_4$ and $C_2N$, with their unique electrical and optical properties, including tunable band gaps and band edge positions, enable effective absorption of visible light and facilitate ideal alignment for photocatalytic reactions [16-18]. Designing heterostructures is thought to be an effective way to boost photocatalytic activity and achieve efficient charge separation. This heterostructure interface extends charge-carrier lifetimes and broadens its light-harvesting spectrum to incorporate visible light [19-21]. The π-conjugated semiconductor with a small band gap of 2.70 eV, graphitic carbon nitride (g-$C_3N_4$), demonstrated comparatively strong and consistent photocatalytic activity with a metal-free property. The g-$C_3N_4$ photocatalyst has been the subject of several studies since the work of Wang et al. Compared to organic π-conjugated semiconductors, graphitic g-$C_3N_4$ often exhibits superior crystallization [22-24]. Its layered structure may make charge transfer more appealing[25]. Additionally, the soft polymer g-$C_3N_4$ is more readily combined with other catalysts than inorganic π-conjugated semiconductors[26]. As a result, the band structure can be readily tuned by doping or by mixing another photocatalyst with g-$C_3N_4$, thereby successfully preventing photo-induced electron-hole recombination. For example, Zhang et al. successfully designed type-II heterojunctions of CdS/g-$C_3N_4$, resulting in enhanced photocatalytic activity for hydrogen photo-generation[27].

All group IV 2D honeycomb materials other than graphene have now been fabricated. Furthermore, chemical functionalization is seen as a viable method for band-gap engineering, with the potential to allow nanoelectronics applications of these exceptional materials [28-30]. Germanium sulphide (GeS), a member of the group IV monochalcogenides, shows a small bandgap and intrinsic ferroelectricity, with effective charge separation and transport properties[31]. However, there are limited works on the direct application of GeS as a photocatalyst. Because designing a heterostructure of g-$C_3N_4$ with the strong polarization of GeS monolayer produces synergistic effects that improve solar energy harvesting, redox capability, and hydrogen production efficiency, making it a better choice for renewable energy applications than single-component materials. These g-$C_3N_4$/GeS heterostructures, however, might exhibit type-II band alignment, built-in electric fields caused by ferroelectric polarization, and suitable band-edge sites for water splitting[32]. Nonetheless, the well-thought-out g-$C_3N_4$/GeS heterostructure offers a promising avenue for the creation of durable, high-efficiency, metal-free photocatalysts for solar-powered hydrogen synthesis. This approach not only enhances solar-to-hydrogen (STH) conversion efficiencies but also paves the way for the exploration and optimization of other 2D van der Waals heterostructures as next-generation photocatalysts for overall water splitting and renewable hydrogen generation.

## 2. COMPUTATIONAL METHODS

First principles Density Functional Theory (DFT) calculations were performed using the projector augmented wave method with plane wave basis set as implemented in Vienna Ab initio Simulation Package (VASP)[33, 34]. The exchange-correlation functional was approximated by using the Perdew–Burke–Ernzerhof (PBE) parametrization of the generalized gradient approximation (GGA)[35]. The free-standing monolayers of g-$C_3N_4$ and GeS and their heterostructures were constructed using VESTA[36]. A vacuum layer of 20 Å was considered along the ‘z’ direction to eliminate the interaction with the periodic layers. We have used the DFT-D3 dispersion corrections as developed by Grimme to account for the long-range van der Waals interaction[37, 38]. The Heyd-Scuseria-Ernzerhof (HSE06) hybrid functional was used to accurately predict the band gaps of the monolayers and heterostructures[39]. The kinetic energy cut-off was set at 520 eV. The convergence criteria for the total energy were set at $10^{-6}$ eV. The Brillouin zone integration was performed using a Γ-centered Monkhorst–Pack[40] grid of $7 \times 7 \times 1$ for geometry optimization, while a $9 \times 9 \times 1$ mesh was employed for the electronic structure calculations. A denser k-mesh of size $18 \times 18 \times 1$ was considered for projected density of states calculations using the tetrahedron method. An ionic relax calculation was performed for both the monolayer and heterostructures to get the optimized structure until the Hellmann-Feynman forces on each atom were less than 0.01 eV/Å. The thermal stability of the monolayer and heterostructure was verified by ab-initio molecular dynamics (AIMD) simulations for a total time of 10 ps in the NVT ensemble using a Nosé-Hoover thermostat. All the computational results were post-processed using the VASPKIT software[41].

## 3. RESULTS AND DISCUSSIONS

### Crystal structure and electronic properties

Figure 1(a,b) indicates the optimized single unit cell of the g-$C_3N_4$ monolayer with a porous layered structure. The optimized lattice constant for the single layer of g-$C_3N_4$ was estimated around 7.15 Å, which is consistent with previous DFT predictions[42, 43] and also with the experimental report[23]. The structure crystallizes in a hexagonal lattice with space group number 187 and space group $P\bar{6}m2$ [44], where the graphitic planes are made up of tri-s-triazine units which are connected by planar amino group[45]. The crystal structure of the GeS monolayer in 2x2x1 supercell form is given in Figure 1 (c,d). The Ge and S atoms are connected to each other at alternate positions, forming a hexagonal buckled lattice with space group $P\bar{3}m1$[46]. After structural relaxation, the optimized lattice constant and Ge-S bond length in a single unit cell of GeS were found to be 3.47 Å and 2.43 Å, respectively, which are in good agreement with literature reports[47-49]. In order to accurately describe the nature of chemical bonding and electron density distribution between the atoms, we have performed the electronic localization function (ELF) calculations[50] for the g-$C_3N_4$ and GeS monolayers (see Figure S1 a,b in the supplementary information (SI) file). The ELF indicates a jellium like homogeneous electron gas and is represented as a contour map normalized to values between zero and one. The region in the plot with ELF values 1.00, 0.50 and 0.00 respectively indicates a fully localized region, a fully delocalized region and very low electron density region respectively. From the Figure S1a, it can be clearly seen that a low charge density can be found at the carbon sites whereas an ELF value close to 1 can be seen near the nitrogen sites close to the hollow region due to the presence of lone pair electrons. However, a low-density region can be seen at the nitrogen atoms forming the C-N rings. The presence of a red region (ELF ~ 1) between the carbon and nitrogen indicates that electrons are fully localized between these atoms, confirming the presence of covalent bonding in the g-$C_3N_4$ monolayer. Similarly, the ELF plot for the GeS monolayer (Figure S1b in the SI) exhibits a high value of ELF at the 'S' sites but a low value of ELF at the 'Ge' sites, which shows that electrons are predominantly localized near the 'S' atoms. This indicates a polar covalent nature of the Ge-S bond with a strong charge transfer from Ge to the S atoms. The heterostructure is constructed by stacking a single unit cell of g-C3N4 with the 2x2x1 supercell of GeS monolayer in order to minimize the lattice mismatch between the two layers (Figure 1e). It was found that the lattice match was around 3.02%, which is within the permitted range (below 5%) for the construction of the heterostructure. The lattice constant of the heterostructure was optimized by applying biaxial strain and was fixed at 7.032Å (Figure S2a in the supporting information file). The interlayer separation between the two layers was optimized with respect to total energy (Figure S2 b in the supporting information file) and was fixed at 3.6 Å (Figure 1f). After structural optimization, the g-$C_3N_4$ monolayer lost its planarity, which is also consistent with previous literature reports (see figure S3 in the SI file). In order to have a better understanding about the crystal structure of individual monolayers and their heterostructure formation, an extended top and side view of the g-$C_3N_4$, GeS monolayer and g-$C_3N_4$/GeS heterostructure is given in supercell form in the figure S4 in the SI file. The thermodynamic stability for the 2D free standing g-$C_3N_4$, GeS monolayers and g-$C_3N_4$/GeS heterostructure was analysed by performing the ab initio molecular dynamics (AIMD) simulations and is illustrated in Figure S5 in the SI file. The calculations were executed at 300 K temperature in two steps. In the first step, the temperature of the system was increased from zero to 300 K up to 10 ps time range using an NVE ensemble with a time step of 1fs. In the second step, the system was kept at 300 K temperature using an NVT ensemble. All the structures show minimal fluctuations in the total energy for the entire range of time indicating a very small structural distortions in both the monolayer and heterostructures. These minor fluctuations in total energy confirm that the monolayers of g-$C_3N_4$, GeS, and their heterostructure are thermodynamically stable.

In this section, we discuss the electronic structure of freestanding g-$C_3N_4$ and GeS monolayer and their heterostructures (Figure 2). The HSE-06 band structure of the pristine g-$C_3N_4$ monolayer indicates an indirect band gap semiconductor with an energy gap of 2.81 eV. The valence band maximum (VBM) and conduction band minimum (CBM) occur at the Γ and K point respectively (Figure 2a). Similarly, the electronic structure of the pristine GeS monolayer exhibits an indirect band gap of 3.31 eV, where the VBM lies along $K - \Gamma$ direction and CBM lies along the $M - \Gamma$ direction respectively (Figure 2b). When a heterostructure is formed by stacking a g-$C_3N_4$ monolayer on top of a GeS monolayer, the band gap becomes 2.17 eV which is less than that of both the monolayers (Figure 2c). The band structure plot indicates an indirect band gap for the heterostructure, where the VBM occurs along the $M - K$ direction and the CBM occurs at the 'M' point of the Brillouin zone. In order to identify the contribution of different orbitals forming the VBM and CBM, we have calculated the projected density of states (PDOS) for both the two free-standing monolayers and their heterostructure. The PDOS calculations for the g-$C_3N_4$ monolayer indicate that the VBM is significantly populated by the '2p' states of nitrogen, whereas the CBM is formed by a mixture of '2p' states of carbon and nitrogen (Figure 2d). Similarly, the PDOS plot for the GeS monolayer clearly illustrates that the VBM is formed with a major contribution from S '3p' states, and the CBM has an almost equal population of Ge '4p' and S '3p' states (Figure 2e). Interestingly, in the heterostructure of g-$C_3N_4$/GeS, it was observed that the VBM is formed by the GeS monolayer with a large population from S '3p' states. On the other hand, the CBM is formed by the '2p' states of carbon and nitrogen with nearly equal population (Figure 2f). Since the VBM and CBM of the heterostructure is formed by two different monolayers, it forms a type-II band alignment which is highly beneficial for photocatalytic water splitting. It is well known that the PBE-GGA functional does not predict the band gap correctly due to derivative discontinuity and self-interaction errors. However, for completeness, we have also mentioned the PBE-GGA functional generated band structure and density of states for the monolayers and heterostructure for comparison (Figure S6, in the SI file).

**Impact of Strain on the electronic structure and energy gap**

Strain is an important parameter in tuning the structural, electronic and optical properties of 2D materials. In the present study, we analyse the impact of biaxial strain on the electronic structure, band gap, band edge positions, optical properties, and HER, OER performance of g-

$C_3N_4$/GeS heterostructure. The biaxial strain was applied by simultaneously compressing/expanding the in-plane lattice parameter of the heterostructure using the relation

$$\varepsilon_a = \frac{a - a_0}{a_0} \times 100\% \quad (1)$$

Where $\varepsilon_a$, $a$ and $a_0$ represent the applied in-plane biaxal strain (in %), the in-plane lattice parameter under strain, and the equilibrium lattice parameter, respectively. The mechanical strain was varied both in compression mode (up to -5%) by decreasing the in-plane lattice parameter and tensile mode (up to +5%) by increasing the lattice parameter, respectively. The variation of band gap for the entire range of strain using PBE and HSE-06 functional is illustrated in Figure 3. The PBE result indicates that the band gap decreases with an increase in tensile strain from zero to +5%. In contrast, with the compression mode, the band gap initially increases from zero to -2% and then decreases systematically up to -5% compressive strain. The use of HSE-06 functional not only brings in a significantly larger magnitude of the band gap but also exhibits a different pattern of variation as compared to the PBE result. The plot indicates that the band gap increases or decreases monotonically from its equilibrium configuration with the application of compressive and tensile strain, respectively (Figure 3). Throughout the entire range of strain applied, the band gap of the heterostructure decreases from 2.8 eV (at -5%) to 1.56 eV (at +5%) which plays a crucial role to enhance the photocatalytic activity under visible light. The different trend in the band gap could be due to the difference in the performance of PBE-GGA and HSE-06 functional towards the exchange-correlation and band edge states. The non-monotonic variation of the band gap in the PBE functional could have arisen from the strain-induced band edge reordering caused by the self-interaction error, which is inherent in the GGA functional. In contrast, the HSE-06 functional stabilizes the electronic states, exhibits a linear variation of band gap with strain and is widely considered to be more reliable. To have a deeper understanding of the band gap and band edges, we performed the strain induced band structure calculations both for the compression and tensile modes (Figure 4). Under -1% compressive biaxial strain, the indirect band gap of the zero-strain structure changes to a direct band gap state (Figure 4a) with the VBM and CBM both at the 'K' point. With an increase in compressive strain from +1% to +5%, the position of VBM almost remains the same, whereas the CBM moves to a higher energy state with respect to the Fermi level (Figure 4 b-e), keeping the direct band gap feature intact. Similarly, when the heterostructure is subjected to tensile biaxial strain, the indirect band gap of the zero strain configuration changes to a direct band gap from +1% to +4% value of $\varepsilon_a$ with both the VBM and CBM at the 'K' point (Figure 4 f-i). At a higher tensile strain of +5%, the heterostructure again transforms to an indirect band gap state with the CBM at the 'K' point and VBM along the $\Gamma - M$ direction (Figure 4j). It was observed that the VBM remains at the 'K' point and the CBM comes closer to the Fermi level as $\varepsilon_a$ changes from +1% to +5% (Figure 4 f-j).

**Strain induced band alignment**

In addition to having a suitable band gap, and type -II band alignment, the photocatalytic activity of two-dimensional van der Waals heterostructures strongly depends on the appropriate band edge position with respect to the vacuum level. For a material to become an efficient photocatalyst, the band edge positions must appropriately align with the water redox potentials, in particular, the hydrogen evolution potential ($H^+/H_2$) and the oxygen evolution potential ($O_2/H_2O$). That means the CBM should lie above the reduction reaction potential, and the VBM should lie below the oxidation reaction potential to carry out the redox reaction. From literature, the standard reduction and oxidation reaction potential of water were found to be -4.44 eV and -5.67 eV, respectively[17]. The calculated band edge position for the unstrained heterostructure is shown in Figure 5a. Using the PBE functional, the band edges were found to be very close to the standard water redox potential. Since the PBE functional underestimates the band gap and band edge position, we have also calculated the band alignment using the HSE-06 functional, which indicates that the CBM lies 0.5 eV above the $H^+/H_2$ reduction potential and VBM lies 0.45 eV below the $O_2/H_2O$ oxidation potential. We note that all these band edge positions were calculated with reference to the vacuum level which was determined using the planar average electrostatic potential. The calculated optical absorption spectrum lies in the visible region, indicating (Figure 5b). The charge density difference also indicates a charge depletion (cyan color) near the g-$C_3N_4$ monolayer and charge accumulation near the GeS monolayer, thus ensuring a charge transfer from g-$C_3$N4 to the GeS monolayer (Figure 5c). Although the VBM and CBM straddle the water redox potential for the unstrained structure, their corresponding positions remain far form the standard reference value. In order to have a more favourable band edge position, we have performed the band alignment calculation under the biaxial strain, which is given in Figure 6. The PBE-GGA functional estimated band edge position was given in Figure 6a for the strain range of -5% to +5% as a reference. The accurate band alignment calculation was executed using the HSE-06 functional is displayed in Figure 6b. From the figure, it was found that with an increase in compressive strain from -1% to -5%, the VBM and CBM lie far from the standard water redox potential. Although the position of VBM does not change appreciably, the CBM exhibits a substantial shift from the -4.44 eV $H^+/H_2$ reduction potential. Similarly, in the tensile biaxial strain region, the CBM remains very close to the water redox potentials up to +3% strain range. In contrast, with further increase in tensile strain up to +5%, the CBM lies below the reduction reaction potential, although the VBM lies below the oxidation reaction potential. This indicates that the heterostructure can not be used for HER when the applied tensile strain is 4% and above. Therefore, the present sample exhibits favourable band edge positions for water splitting, HER and OER for $\varepsilon_a = +1\%, +2\%$ and $+3\%$.

**HER Activity with Strain**

To further verify the validity of the band alignment result and the kinetic feasibility of photocatalytic water splitting, we have analysed the hydrogen evolution reaction (HER) performance of the g-$C_3N_4$/GeS heterostructure under a tensile strain range from 1% to 3%. The production of hydrogen via water splitting involves two reaction steps. The first step is the adsorption of atomic hydrogen on the surface of the heterostructure at suitable adsorption sites. The second step is the formation and release of molecular hydrogen. The adsorption of atomic hydrogen occurs through the Volmer reaction step, while the interaction with another hydrogen is followed by the subsequent evolution of a hydrogen molecule either via the Heyrovsky or Tafel reaction pathway. Before proceeding with the HER analysis, we have calculated the adsorption energy of hydrogen ($E_{ads}$) using the relation

$$E_{ads} = E_{surface+H} - E_{surface} - \frac{1}{2} E_{H_2} \quad (2)$$

Where, $E_{surface}$, $E_{surface+H}$, $E_{H_2}$ represents the total energy of a clean, adsorbed surface with a hydrogen atom and a hydrogen molecule respectively. The HER performance was analysed by calculating the Gibbs' free energy change ($\Delta G_{HER}$) for hydrogen adsorption, which serves as a crucial descriptor for the HER. $\Delta G_{HER}$ was calculated using the relation

$$\Delta G_{HER} = \Delta E_{ads} + \Delta E_{ZPE} - T\Delta S_H \quad (3)$$

Where $\Delta E_{ads}$ is the hydrogen adsorption energy, $\Delta E_{ZPE}$ and $\Delta S_H$ are respectively the change in zero point energy and entropy difference between the adsorbed hydrogen and gas phase hydrogen, T is the temperature (298.15 K). The entropy of the gas-phase hydrogen molecule was derived from the U.S. National Institute of Standards and Technology (NIST) database.

For an efficient photocatalyst, the Gibbs' free energy change should be neither too strong nor too weak, as it determines the strength of the adsorbing hydrogen. If the adsorption is too strong then it will exhibit resistance to the desorption of the hydrogen molecule. Similarly, weaker Gibbs' free adsorption energy indicates that the protons can not be easily attracted towards the catalyst surface. According to the Sabatier principle, for an ideal catalyst the Gibbs' free energy chang should be very close to zero ($\Delta G_{HER} \approx 0$) in order to have maximum reaction rate.

A single hydrogen atom was placed at 2Å height from the sample surface at different possible adsorption sites, and an ionic relax calculation was performed for each configuration. The structure with minimum energy is considered as the ground state, which is used for the next calculation. The calculations were performed both for the unstrained structure and for biaxial tensile strain range up to +3%. The Gibbs' free energy change for all these configurations was calculated using equation (3) and is displayed in Figure 7. It was found that the reference heterostructure exhibits an overpotential of around -0.20 eV. With an increase in 1% tensile biaxial strain, the $\Delta G_{HER}$ becomes -0.1 eV, indicating an enhanced HER activity. Similarly, with 2% tensile biaxial strain, the $\Delta G_{HER}$ becomes slightly positive (+0.1 eV), which is still close to zero, indicating extremely good HER performance. At a larger positive tensile strain of 3%, the Gibbs' free energy change increases to +0.38 eV, indicating a moderate activity for HER. These observations are consistent with the band edge calculations results given in Figure 6. Since at a larger value of tensile strain, the CBM comes below the hydrogen reduction potential, we have not considered the impact of higher strain on the HER overpotential study. From the above discussion, we found clear evidence that the present heterostructure exhibits enhanced HER activity with low tensile biaxial strain up to 3%.

**OER Activity with Strain**

In this section, we analyse the oxygen evolution reaction activity of the g-$C_3N_4$/GeS heterostructure under mechanical strain to get a better understanding of water splitting. The mechanism of OER involves the following four sequential one-electron transfer reaction steps which are as follows.

$$H_2O\ (l) + * \xrightarrow{\Delta G_1} OH^* + H^+ + e^- \tag{4}$$

$$OH^* \xrightarrow{\Delta G_2} O^* + H^+ + e^- \tag{5}$$

$$O^* + \; H_2O\,(l) \xrightarrow{\Delta G_3} OOH^* + H^+ + e^- \tag{6}$$

$$OOH^* \xrightarrow{\Delta G_4} * + O_2(g) + H^+ + e^- \tag{7}$$

Where (l) refers to the liquid phase and (g) represents the gas phase, * stands for the active sites, $O^*$, $OH^*$, and $OOH^*$ indicates the intermediate configuration adsorbed with ions. Here $\Delta G_1, \Delta G_2, \Delta G_3, \Delta G_4$ represent the calculated Gibbs' free energy change for the $OH^*$, $O^*$ and $OOH^*$ intermediates and the final structure respectively. The free energy was calculated using the relation

$$\Delta G_{OER} = \Delta E + \Delta E_{ZPE} - T\Delta S + \Delta G_U + \Delta G_{PH} \tag{8}$$

Where $\Delta E$ is the reaction energy calculated using DFT calculations, $\Delta E_{ZPE}$ is the zero point energy, $\Delta S$ is the change in entropy. The temperature was set at 298.15 K. Since the calculations were performed at PH=0 in the present study, the terms $\Delta G_U$ and $\Delta G_{PH}$ vanish in the expression 8. The OER overpotential was calculated using the method developed by Nørskov et al. [51] and can be expressed as

$$\eta = \frac{\max\{\Delta G_1, \Delta G_2, \Delta G_3, \Delta G_4\}}{e} - 1.23\,V \tag{9}$$

The calculated free energy profile for the oxygen evolution reaction for the g-$C_3N_4$/GeS heterostructure is given in Figure 8. The calculated values of $\Delta G_1, \Delta G_2, \Delta G_3, \Delta G_4$ for the reference and strained structures are mentioned in the corresponding figures. For the reference structure (zero strain), the overpotential (η) and onset potential were estimated to be 2.17 V and 3.4 eV respectively (Figure 8a) indicating its poor catalytic activity. With increase in tensile strain from 0% to 1%, the OER overpotential indicates a substantial decrease to 1.06 V. Similarly, the onset potential also decreases from 3.4 eV to 2.29 eV (Figure 8b). As the tensile strain increases to 2%, the onset potential and overpotential further decreases to 2.25 V and 1.02 V respectively (Figure 8c). At higher tensile strain at 3%, the onset potential and η value reduces to 2.20 V and 0.97 V respectively (Figure 8d). The systematic reduction in onset potential and over potential with applied tensile strain indicates the significant improvement of the OER activity of the heterostructure, which originates from the optimized adsorption strength of the oxygen intermediates and thereby facilitates the OER process. We would like to mention that in all the strained configurations and the reference structure, the highest value of $\Delta G$ occurs in the third intermediate step (OOH* derivative) of the OER reaction which serves as the rate-determining step. The OER energy profile analysis is also consistent with the band alignment result with strain (Figure 5b). As the tensile strain increases the VBM remains far from the oxidation potential of $H_2O/O_2$ and does not shift significantly towards the stand value (-5.67 eV) under strain. indicating that the photogenerated holes carry limited oxidation power to drive the oxygen evolution reaction which is reflected due to the presence of large OER overpotential. Although, the application of tensile strain reduces the OER overpotential from 2.17 V at 0% strain to 0.97 V at 3% strain, the OER still requires a considerable amount of energy due to the limited oxidation capability associated with the position of VBM. Overall, the band alignment, HER and OER analysis with strain suggest that the present heterostructure exhibits more favourable condition for HER than OER, and the application of strain partially enhances the OER potential by modification in the surface electronic structure making the system moderately active for OER.

**Optical Properties**

In addition to the oxidation and reduction capabilities, the optical properties of the 2D materials plays an important role in estimating the efficiency of the photocatalyst for water splitting. Since the photocatalyst gets activated upon interaction with photon energy, it is equally important to verify whether the proposed material under study can absorb a maximum fraction of visible light, which constitutes around 42% of the solar spectrum. The calculated optical absorption spectra of the g-$C_3N_4$/GeS heterostructure are shown in Figure S7 in the SI file as a function of photon energy. From the figure, it is evident that the optical absorption of the heterostructure increases with photon energy in the visible region and becomes maximum at 3.56 eV. This maximum absorption could be due to the $n \rightarrow \pi^*$ electron transition occurring due to the presence of a lone pair of electrons on the nitrogen atoms. Similar observations are also reported in the literature for other 2D systems such as CdSe/$C_2N$ and CdS/$C_2N$ heterostructures[52]. As the system is subjected to biaxial compressive (-5%) and tensile (+5%) strain, the optical absorption edge lies within the visible region. Interestingly the absorption peak at 3.56 eV for the reference structure exhibits a substantial shift to 3.13 eV and lies in the visible region at +5% tensile strain. This red shift of the optical absorption is consistent with the band gap reduction result of the heterostructure at larger tensile strain (Figure 3). The above observations clearly indicates that the present g-$C_3N_4$/GeS heterostructure can absorb the maximum portion of visible light which can be further enhanced with tensile strain to exhibit improved photocatalytic water splitting activity for hydrogen generation.

**Photocatalytic water splitting mechanism**

To better understand the charge carrier distribution between the two monolayers forming the heterostructure, we compute the plane-averaged differential charge density ($\Delta\rho$), using the formula

$$\Delta\rho = \rho_{heterostructure} - \rho_{monolayer1} - \rho_{monolayer2} \quad (10)$$

Where $\rho_{heterostructure}$ denotes the charge density of heterostructure g-$C_3N_4$/GeS, $\rho_{monolayer1}$denotes the differential charge density for g-$C_3N_4$, while $\rho_{monolayer2}$ denotes the differential charge density for GeS.

The Bader analysis shows that the lower layer (GeS) loses charge +0.056e, while the upper layer (g-$C_3N_4$) gains a charge of -0.056e. Therefore, charge transfer direction is from the Ges monolayer to the g-$C_3N_4$ monolayer.

Since the Fermi energy levels of the isolated materials are different from each other. The g-$C_3N_4$ layer has a lower Fermi energy than GeS monolayer. Upon the formation of the heterostructure, the Fermi energy tends to equalize through the flow of electrons from the GeS material, with a higher Fermi energy, to the g-$C_3N_4$ material, with a lower Fermi energy, until the Fermi energy is balanced. At this point, a potential difference will be generated by the internal electric field at the contact interface. A built-in electric field of 2.88 eV is directed from g-$C_3N_4$ to GeS. The work function that tells about the capacity to bind surface electrons and affects the internal electric field, as well as interfacial electron transport, is calculated as follows-

$$\Phi = E_{vacuum} - E_F \quad (11)$$

Where $E_{vac}$ denotes the vacuum level, and $E_f$ defines the Fermi level. The obtained value of Φ for g-$C_3N_4$/GeS heterostructure is 5.674 eV (see Figure 9a). Figure 9b describes the complete mechanism of photocatalytic water splitting. When electron transfer takes place between GeS, and g-$C_3N_4$, the energy bands of GeS bend upward while the bands of g-$C_3N_4$ bend downward. This bending, combined with a built-in electric field (2.88eV) directed from g-$C_3N_4$ to GeS, establishes energy barriers that influence charge propagation. There are three charge-carrier paths to transport 1. Interlayer coupling among the VBM of GeS and the CBM of g-$C_3N_4$, where the recombination of electrons and holes takes place. 2. The movement of electrons from the CBM of GeS and the CBM of g-$C_3N_4$, and the holes from the VBM of g-$C_3N_4$ to the VBM of GeS. 3. Direct charge transfer between the CBM of the GeS layer and the VBM of the g-$C_3N_4$ layer. The potential barrier formed by band bending and the built-in electric field restricts the amount of photogenerated charge carriers that can migrate along paths 2 and 3. While paths 2 and 3 are restricted, the band bending increases the rate of recombination of photogenerated charge carriers in the intermediate layer via path 1. This specific carrier movement leads to the accumulation of specific charges on separate layers, creating an ideal site for water splitting. Electrons accumulate in the conduction band (CB) of g-$C_3N_4$, giving a necessary site for the reduction of water (producing hydrogen), and holes accumulate in the valence band (VB) of GeS, giving a necessary site for the oxidation of water (producing oxygen). By ensuring the oxygen evolution reaction (OER) and hydrogen evolution reaction (HER) taking place on distinct semiconductor layers, this spatial separation improves overall photocatalytic performance and prevents instant recombination.

## 4. CONCLUSION

In this work, we present a comprehensive analysis of the photocatalytic activity of g-$C_3N_4$/GeS heterostructure for water splitting using first principles DFT calculations. While the monolayer of g-$C_3N_4$ and GeS possesses a large indirect band gap, their heterostructure exhibits a reduced energy gap of 2.17 eV. The projected density of states calculations confirms the presence of type-II band alignment in the heterostructure which is favourable for charge carrier separation with reduced recombination. The strain induced modification in the electronic structure calculation indicates an indirect to direct band gap transition with the CBM slowly approaching towards the Fermi level with an increase in tensile strain. The band edge position calculations indicates that both the VBM and CBM straddle the standard water redox potential satisfying the condition for water splitting. To validate the photocatalytic activity further, HER and OER analysis were executed both for the reference and strained structure. The Gibbs' free energy of the heterostructure at zero strain was around 0.2 eV which gradually becomes close to zero with an increase in tensile strain indicating its excellent HER capability. However, the OER analysis for the reference heterostructure exhibits a large overpotential around 2.17 V which is reduced to 0.97 V with 3% tensile strain indicating enhancement of OER performance. The optical adsorption spectra indicates that the absorption edge falls in the visible region within -5% to +5% biaxial strain range thus indicating the suitability for visible light water splitting. The present work can be useful to the experimentalist in designing artificial strained g-$C_3N_4$/GeS heterostructure for the optimal performance in photocatalytic water splitting and shed light on the technological application for sustainable hydrogen production.

## ACKNOWLEDGMENTS

The authors would like to acknowledge the use of high-performance computing facility at National Institute of Science Education and Research (NISER).

## DATA AVAILABILITY STATEMENT

The data that support the findings of this study are available from the corresponding author upon reasonable request.

## FUNDING

The authors received no external funding for this research work.

## CONFLICT OF INTEREST

The authors have no conflicts to disclose.

## SUPPORTING INFORMATION

The additional data that support the findings of this article are available in the Supplementary Information.

## Figures

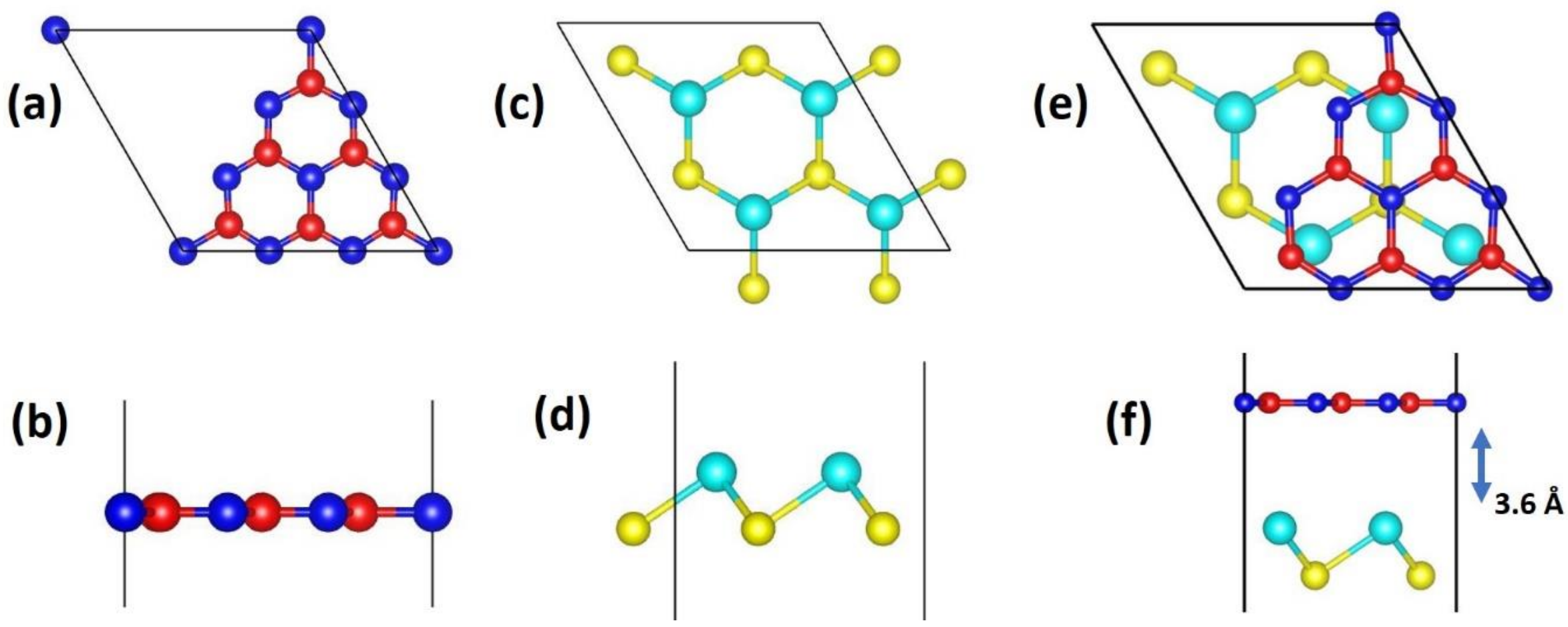

**Figure 1:** Schematic representation of (a,b) g-$C_3N_4$, (c,d) GeS free standing monolayers and (e,f) g-$C_3N_4$/GeS van der Waals heterostructure. The top (bottom) panel indicates the top (side) view of the corresponding unit cells. The colour codes for the atoms are as follows: blue (nitrogen), red (carbon), yellow (sulfur) and cyan (germanium).

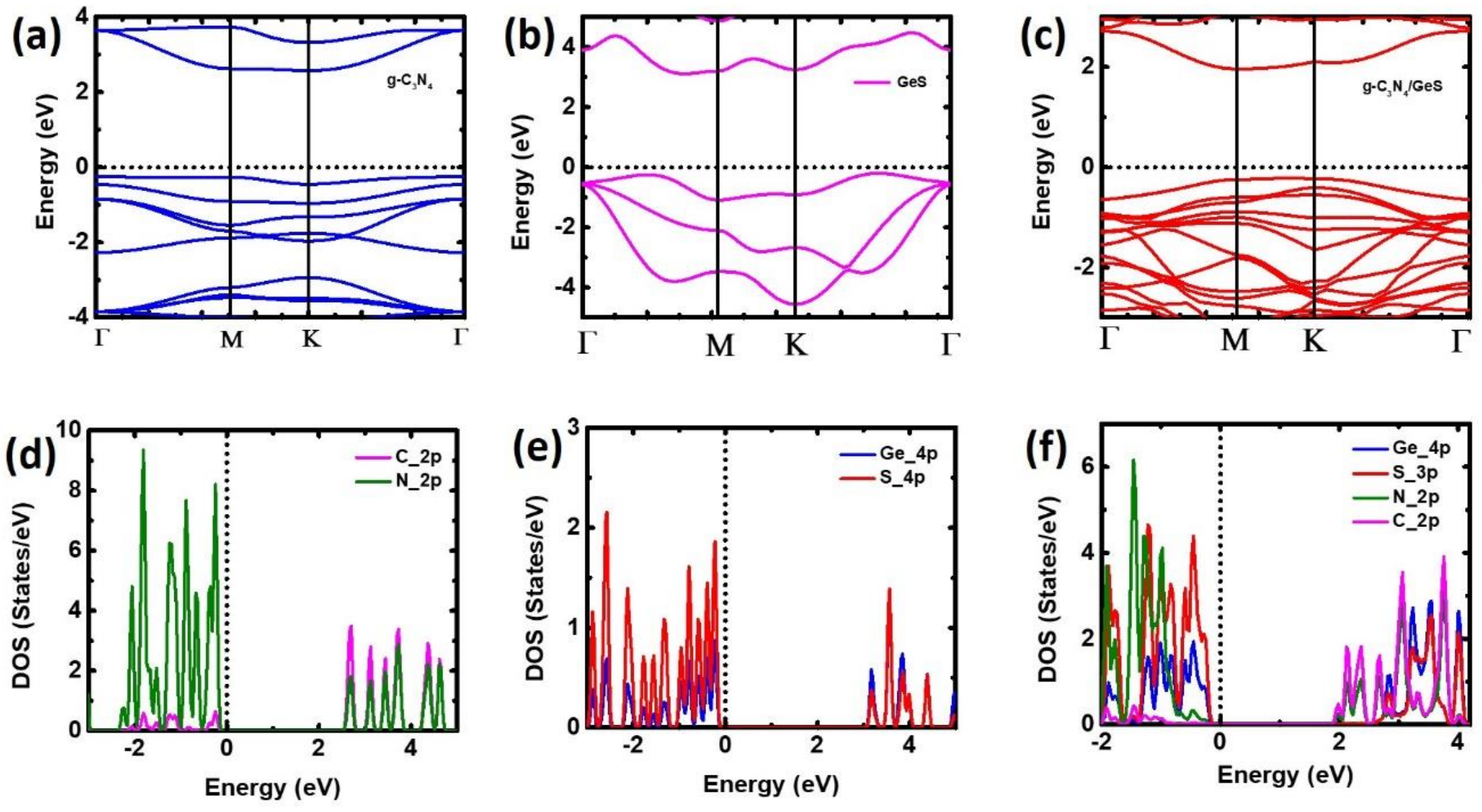

**Figure 2:** Electronic band structure and projected density of states of (a,d) g-$C_3N_4$ (b,e) Ges monolayers and (c,f) g-$C_3N_4$/GeS van der Waals heterostructure. The dotted line indicates the Fermi level, which is set to zero. All the calculations were performed using HSE-06 functional.

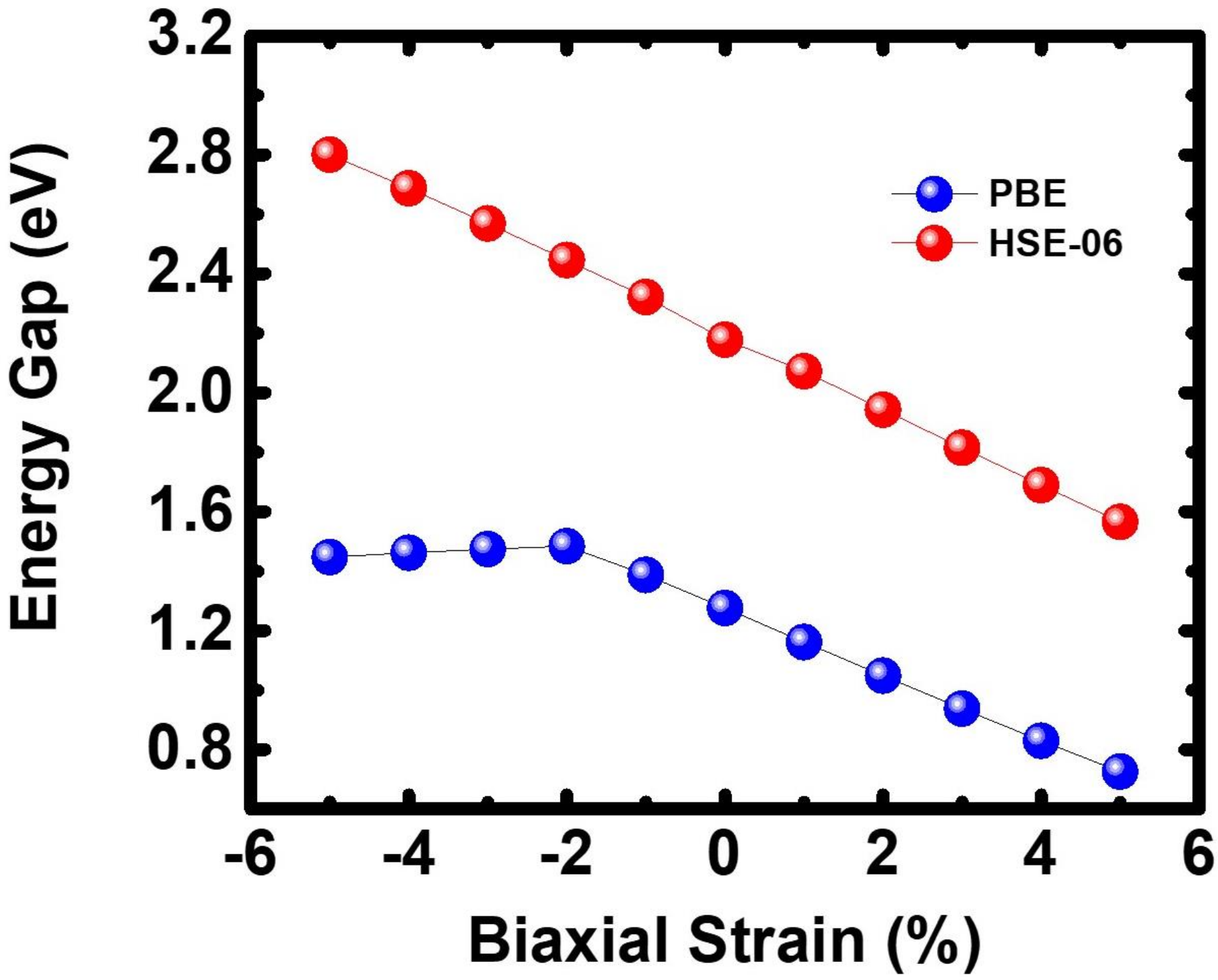

Figure 3: Electronic band gap variation in g-$C_3N_4$/GeS van der Waals heterostructure as a function of applied biaxial strain using PBE-GGA and HSE-06 functional.

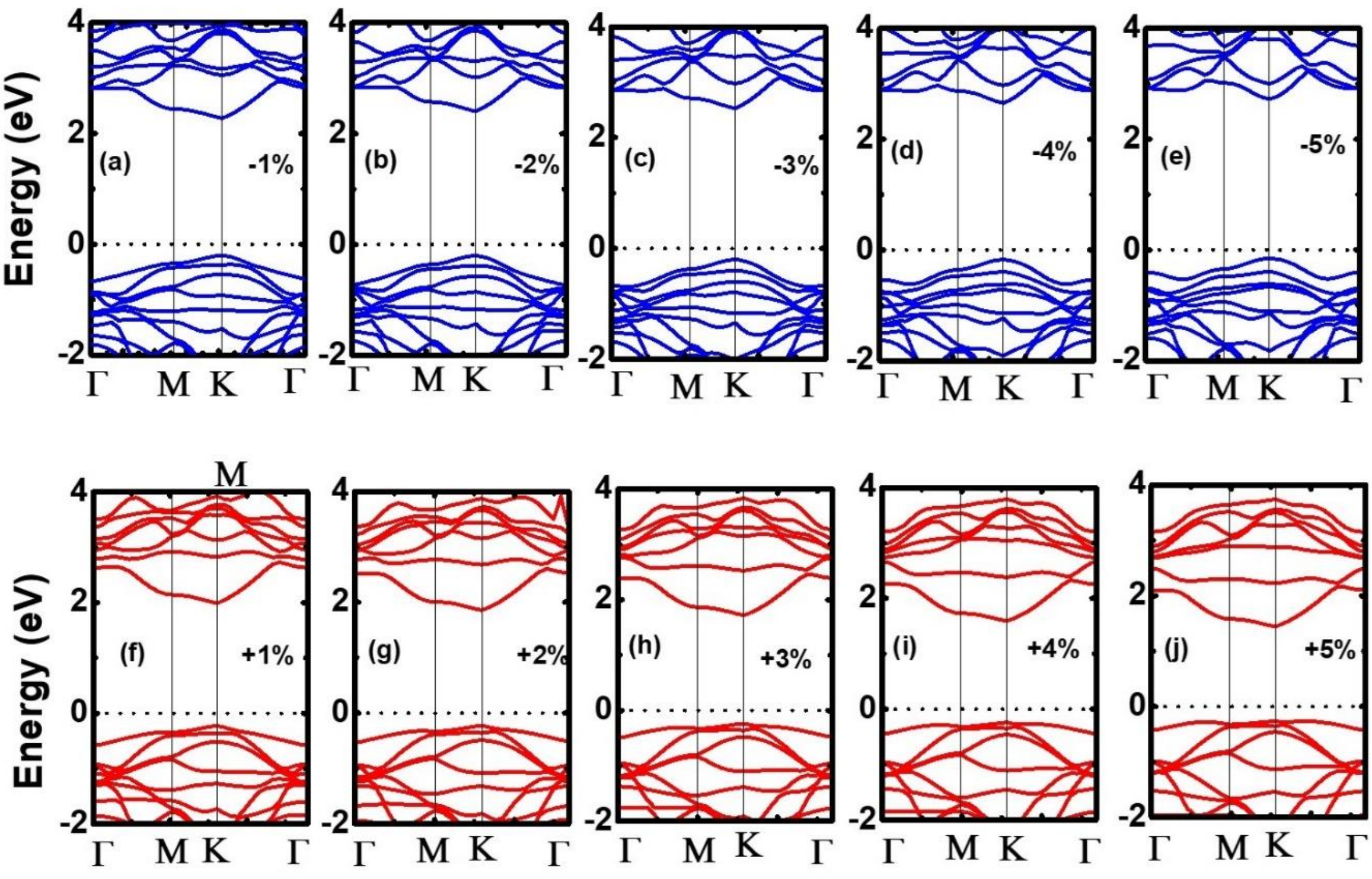

Figure 4: Band structure evolution of g-$C_3N_4$/GeS van der Waals heterostructure as a function of applied biaxial strain using HSE-06 functional.

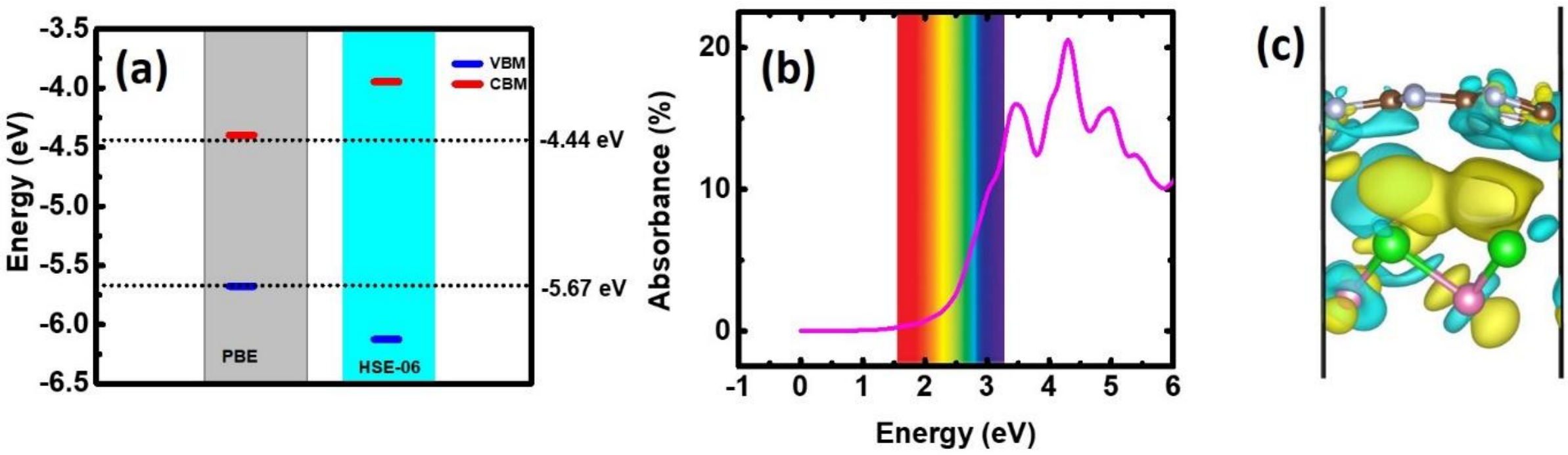

Figure 5: (a) Band edge positions with respect to vacuum level, in unstrained g-$C_3N_4$/GeS van der Waals heterostructure, calculated using PBE-GGA and HSE-06 functionals. The dotted horizontal lines indicate the standard oxidation and reduction potential of water, respectively. (b) optical absorption spectra and (c) charge density difference (iso-surface value = 0.0015) in unstrained g-$C_3N_4$/GeS heterostructure.

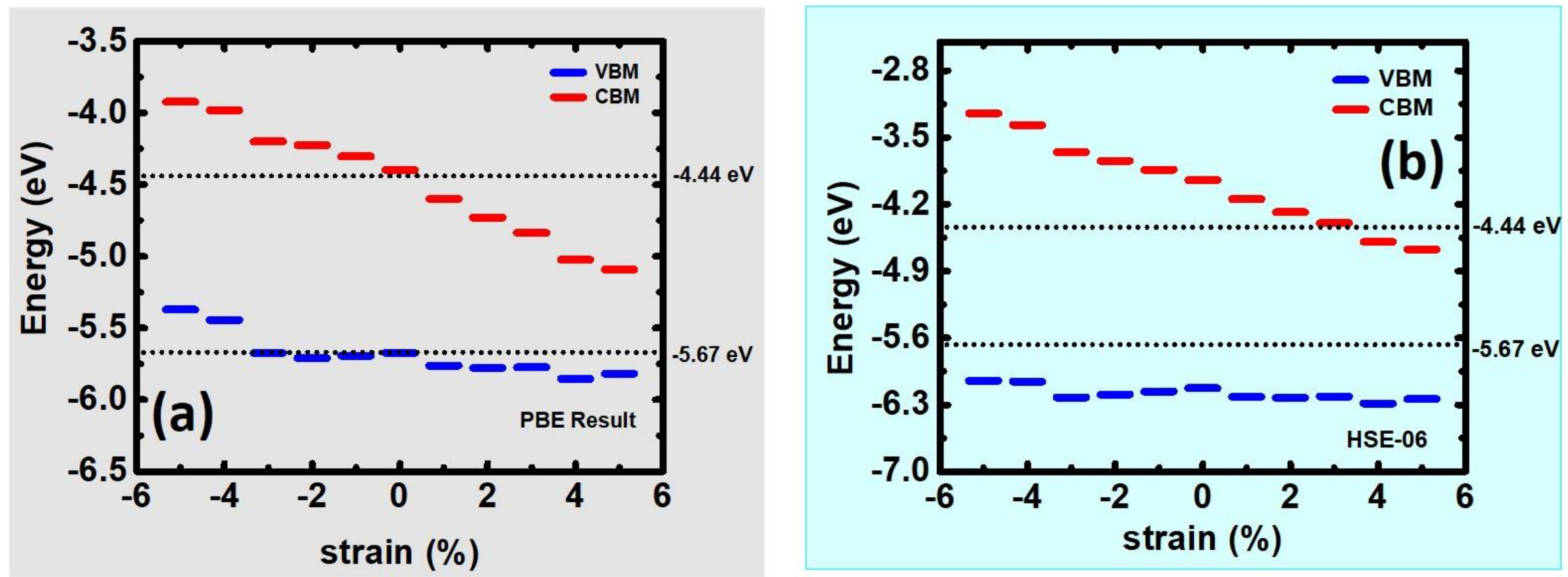

Figure 6: Strain-induced band edge positions with respect to vacuum level in g-$C_3N_4$/GeS van der Waals heterostructure, calculated using (a) PBE-GGA and (b) HSE-06 functionals. The dotted horizontal lines indicate the standard oxidation and reduction potential, respectively.

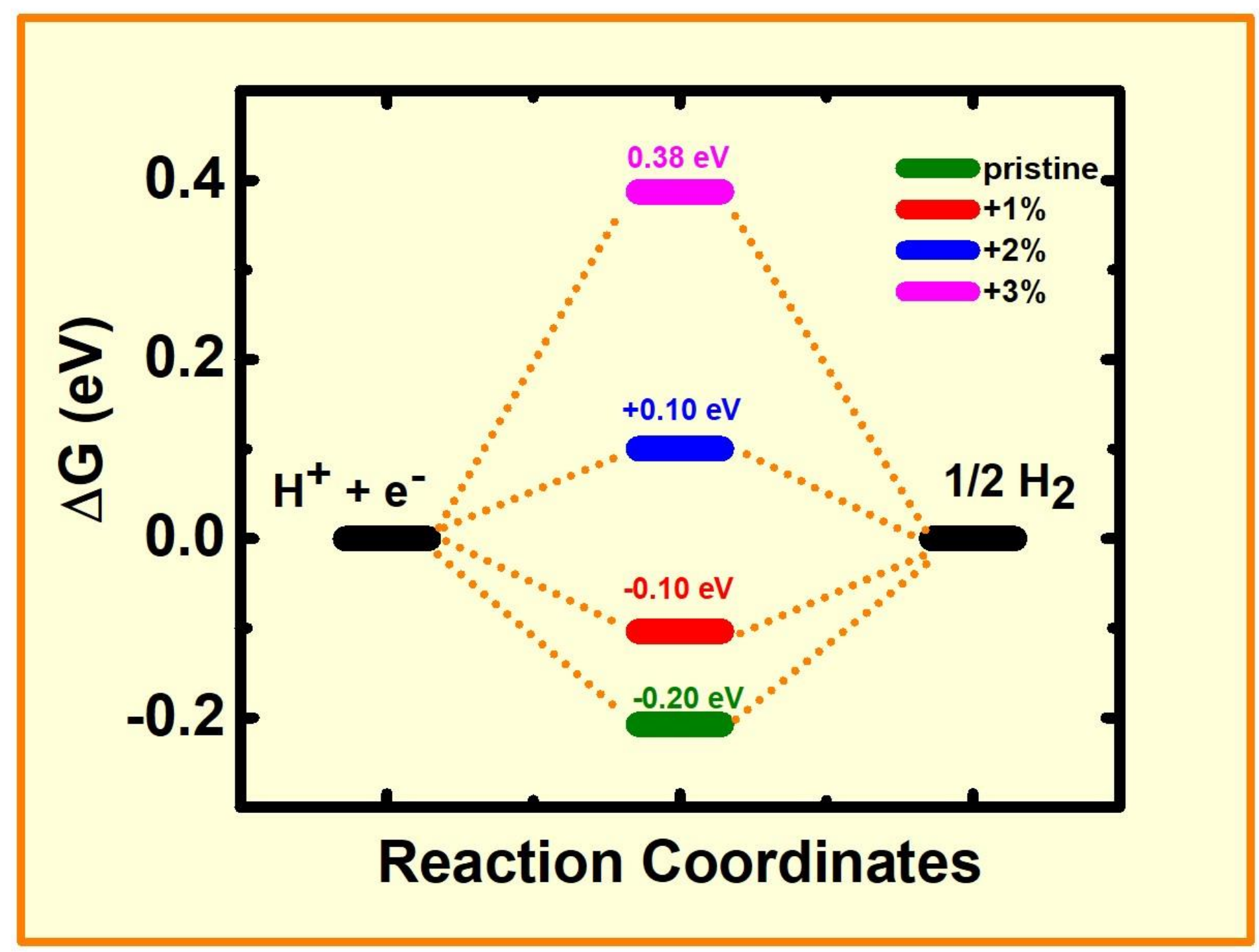

Figure 7: Gibbs' free energy change for g-$C_3N_4$/GeS van der Waals heterostructure with tensile biaxial strain during hydrogen evolution reaction (HER) on the surface.

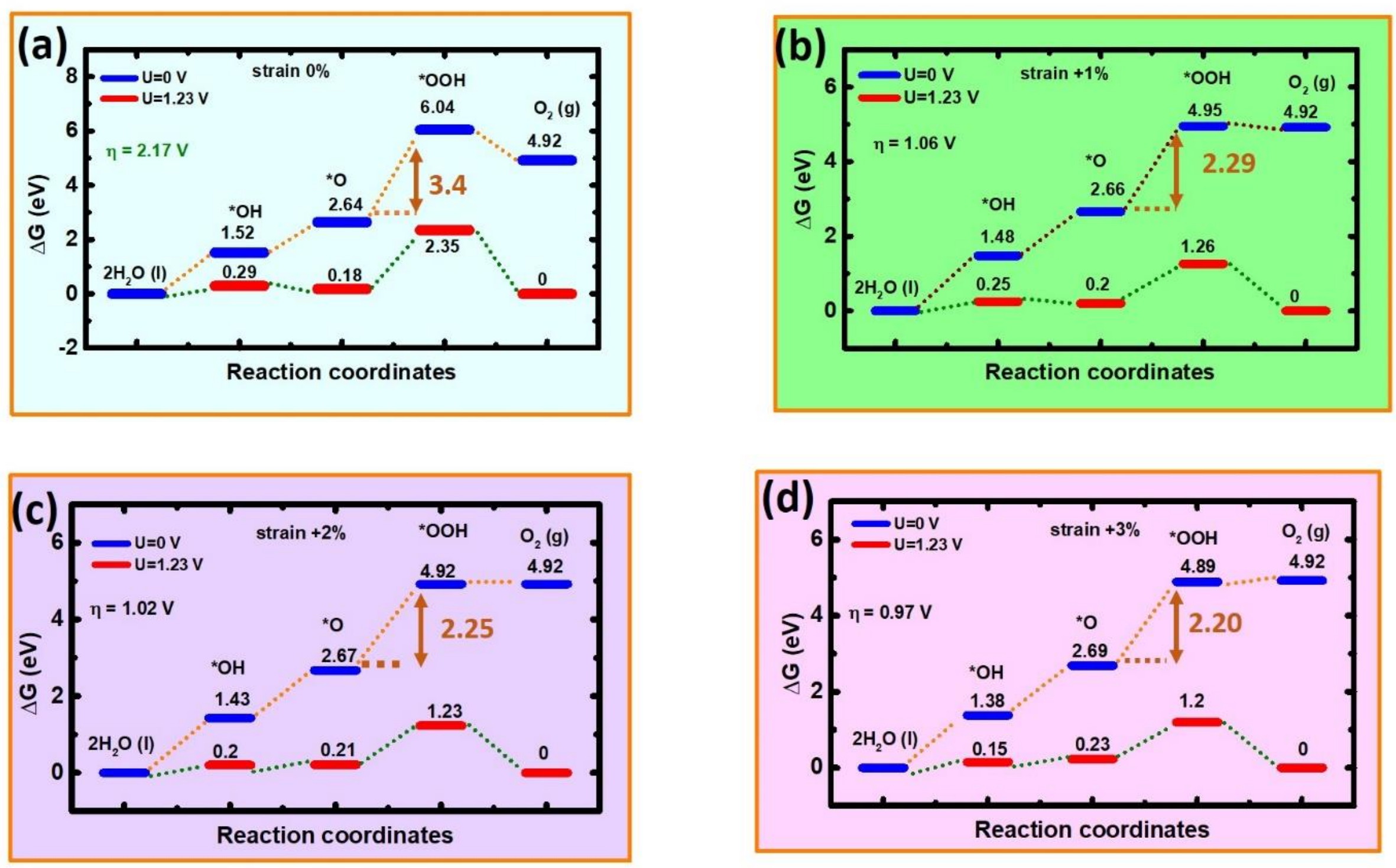

Figure 8: Gibbs' free energy change plot for g-$C_3N_4$/GeS van der Waals heterostructure with tensile biaxial strain during oxygen evolution reaction (OER) on the surface.

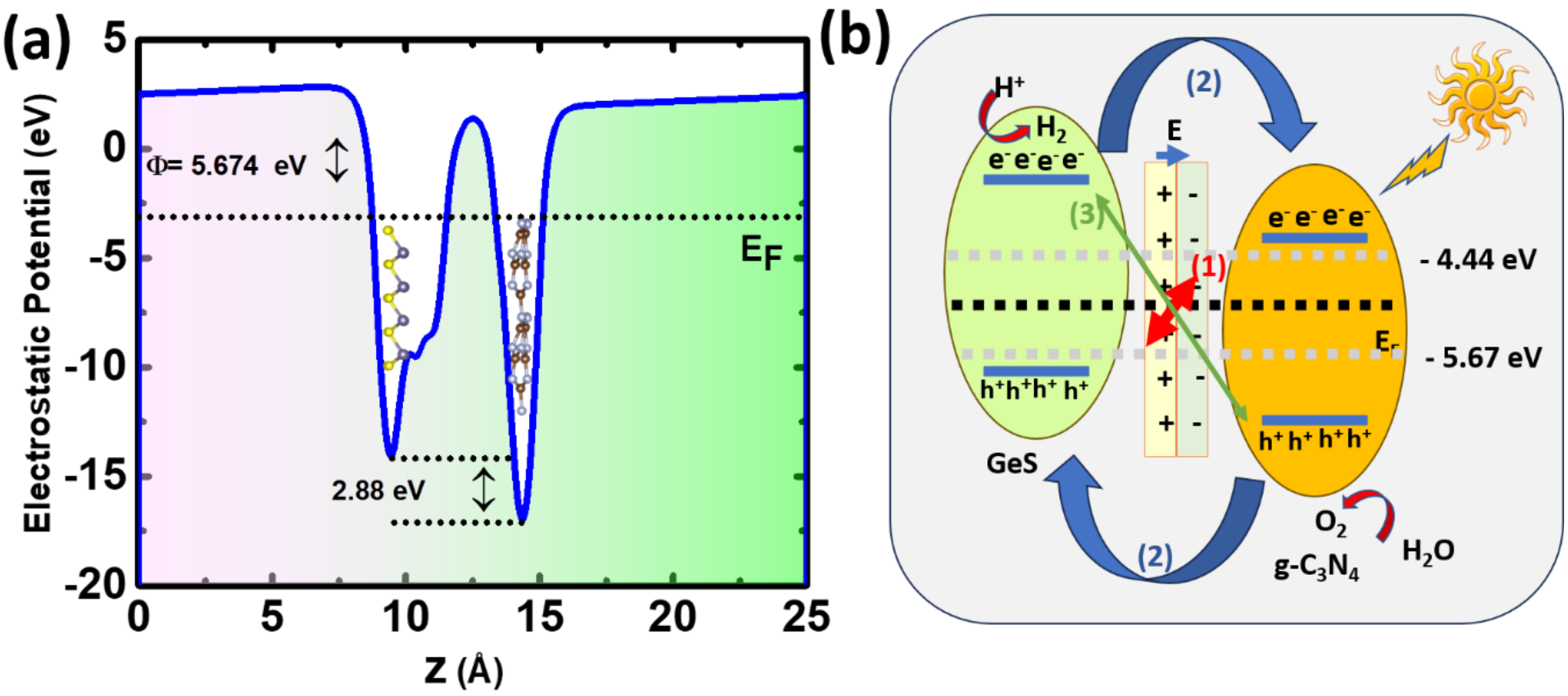

Figure 9: (a) Work function, built-in electric field map and (b) photocatalytic water splitting reaction mechanism in g-$C_3N_4$/GeS van der Waals heterostructure

# Supplementary Information

## Strain-Driven Electronic and Catalytic Modulation of g-$C_3N_4$/GeS van der Waals heterostructure for Photocatalytic Water Splitting

Soumendra Kumar Das[1], Smruti Ranjan Parida[1], Tapas Kumbhakar[1], Prasanjit Samal[2] and Sridhar Sahu[1*]

*[1] Department of Physics, Indian Institute of Technology (Indian School of Mines) Dhanbad, Dhanbad-826004, Jharkhand, India*

*[2]School of Physical Sciences, National Institute of Science Education and Research (NISER) Bhubaneswar, HBNI, Jatni, Khurda-752050, Odisha, India*

**corresponding authors: e-mail: sridharsahu@iitism.ac.in*

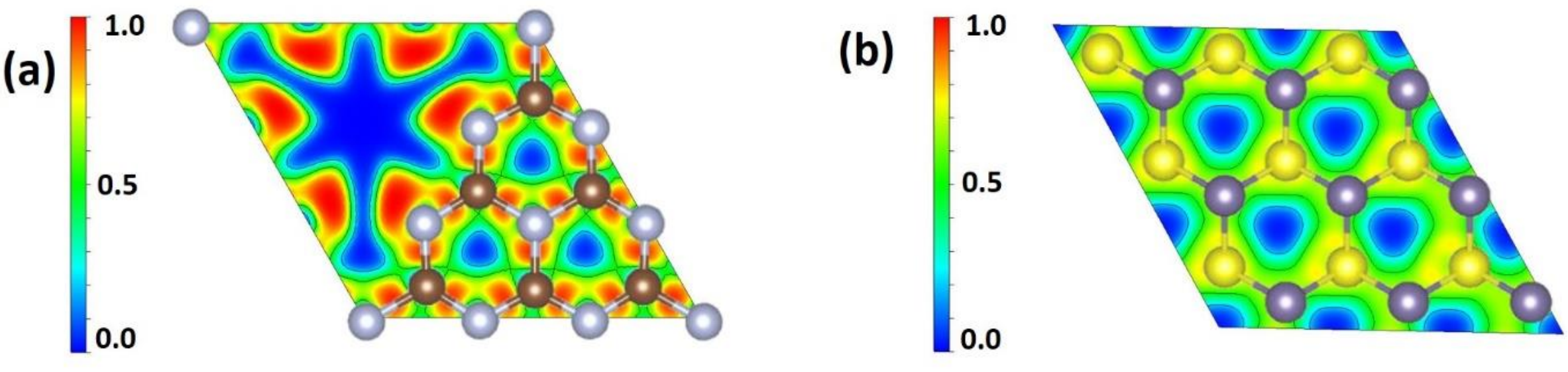

Figure S1: Electron Localization Function (ELF) of free standing (a) g-$C_3N_4$ and (b) GeS monolayer viewed from the [001] direction.

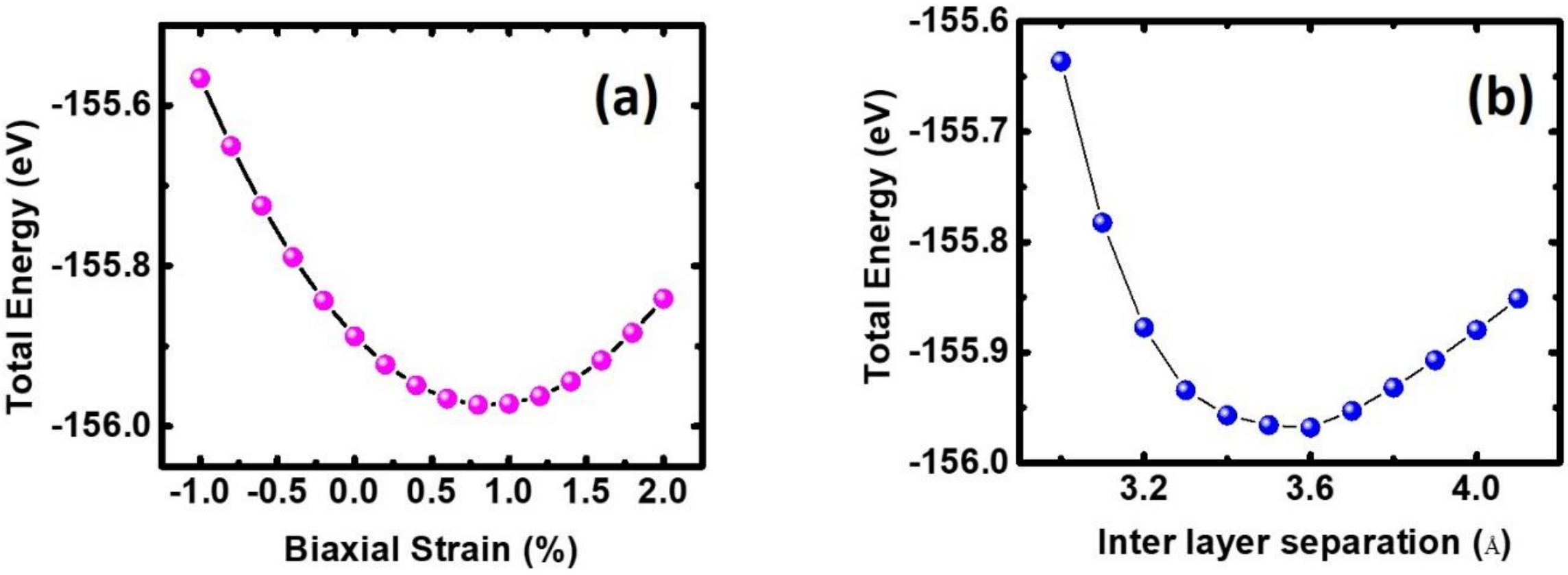

Figure S2: Total energy optimization of g-$C_3N_4$/GeS heterostructure with respect to (a) biaxial strain and (b) inter layer separation.

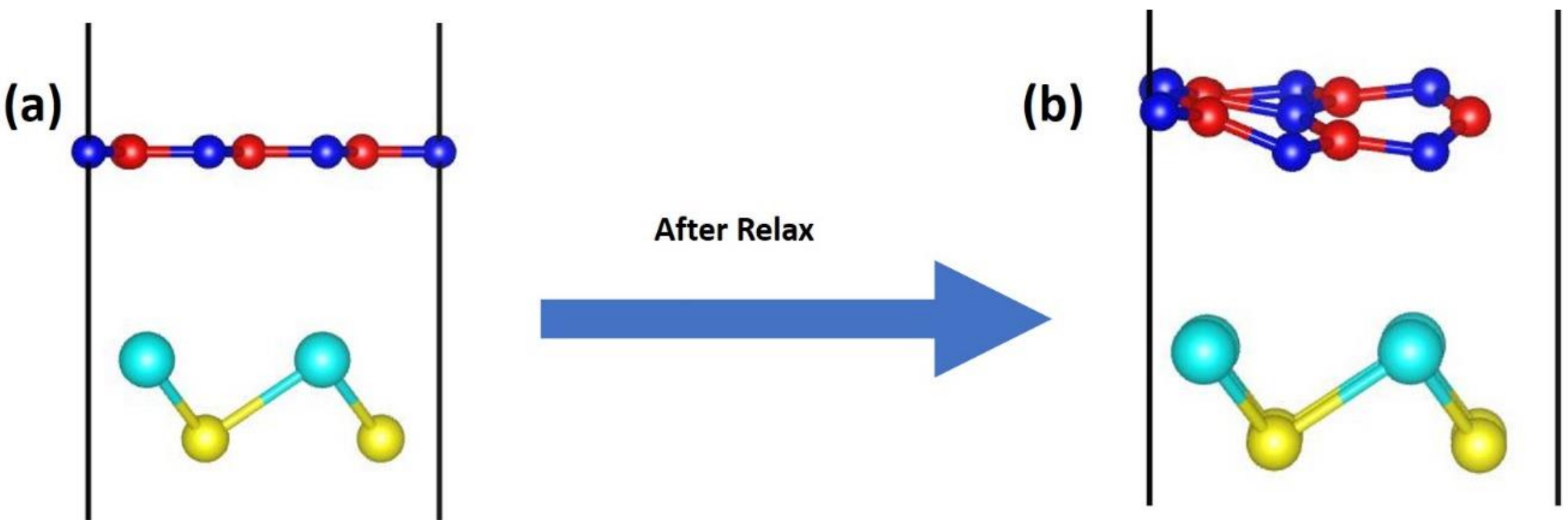

Figure S3: Side view of g-$C_3N_4$/GeS heterostructure before and after geometrical optimization.

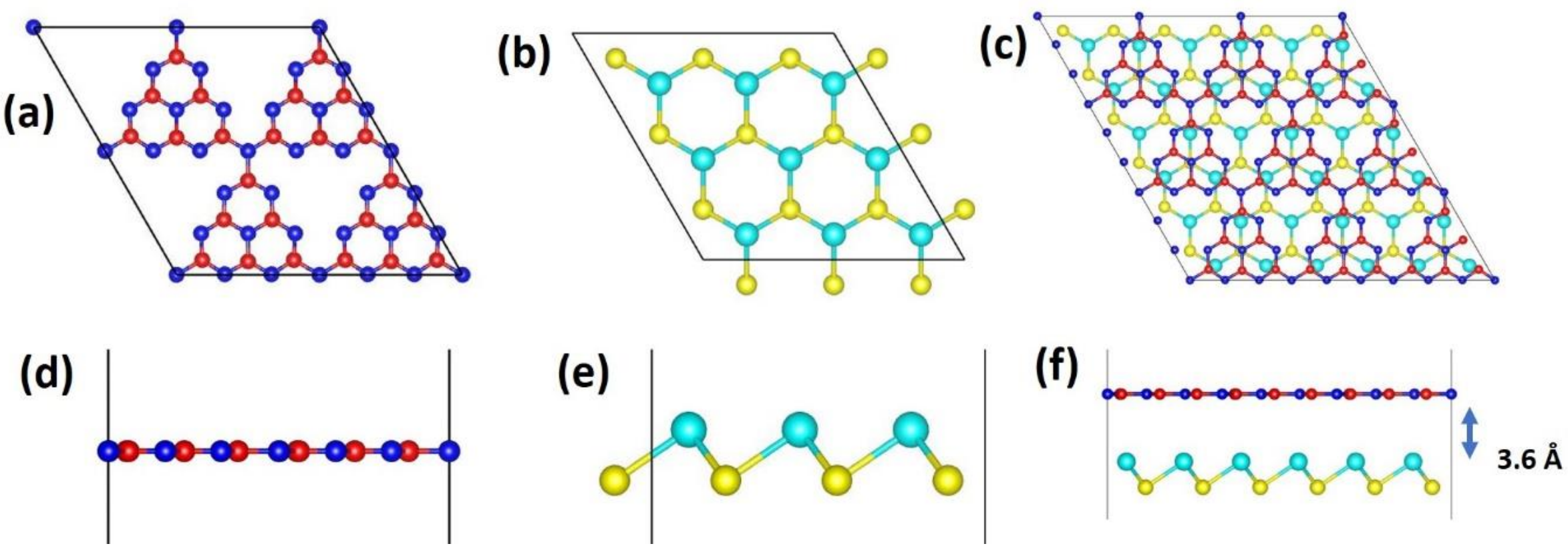

Figure S4: Extended top and side view of (a,d) g-$C_3N_4$ monolayer (b,e) GeS monolayer and (c,f) g-$C_3N_4$/GeS heterostructure.

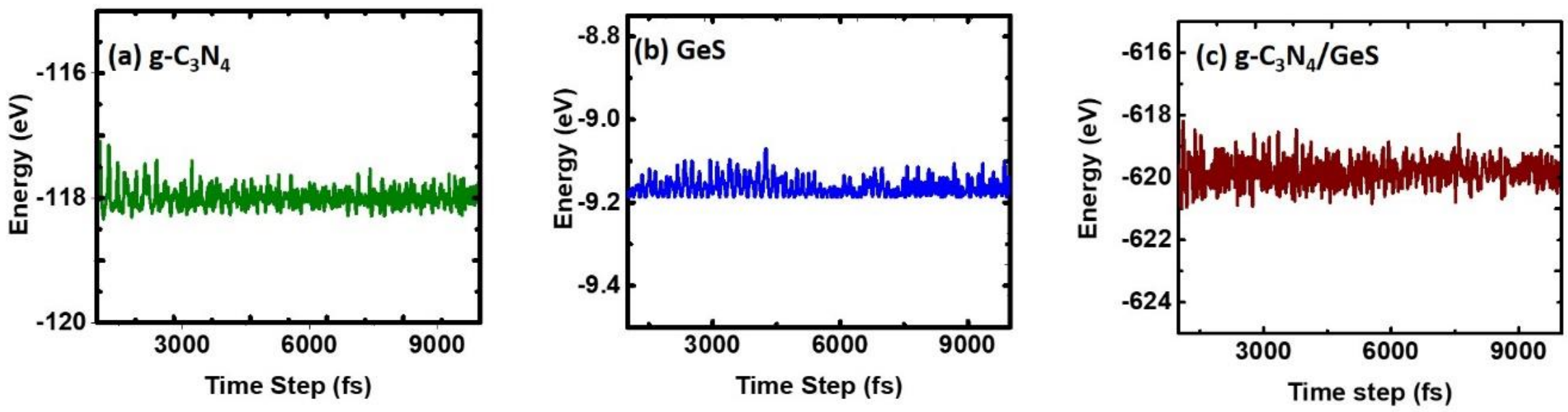

Figure S5: AIMD simulations: Energy fluctuation as a function of time step for (a) g-$C_3N_4$ (b) GeS monolayers and (c) g-$C_3N_4$/GeS heterostructure, respectively.

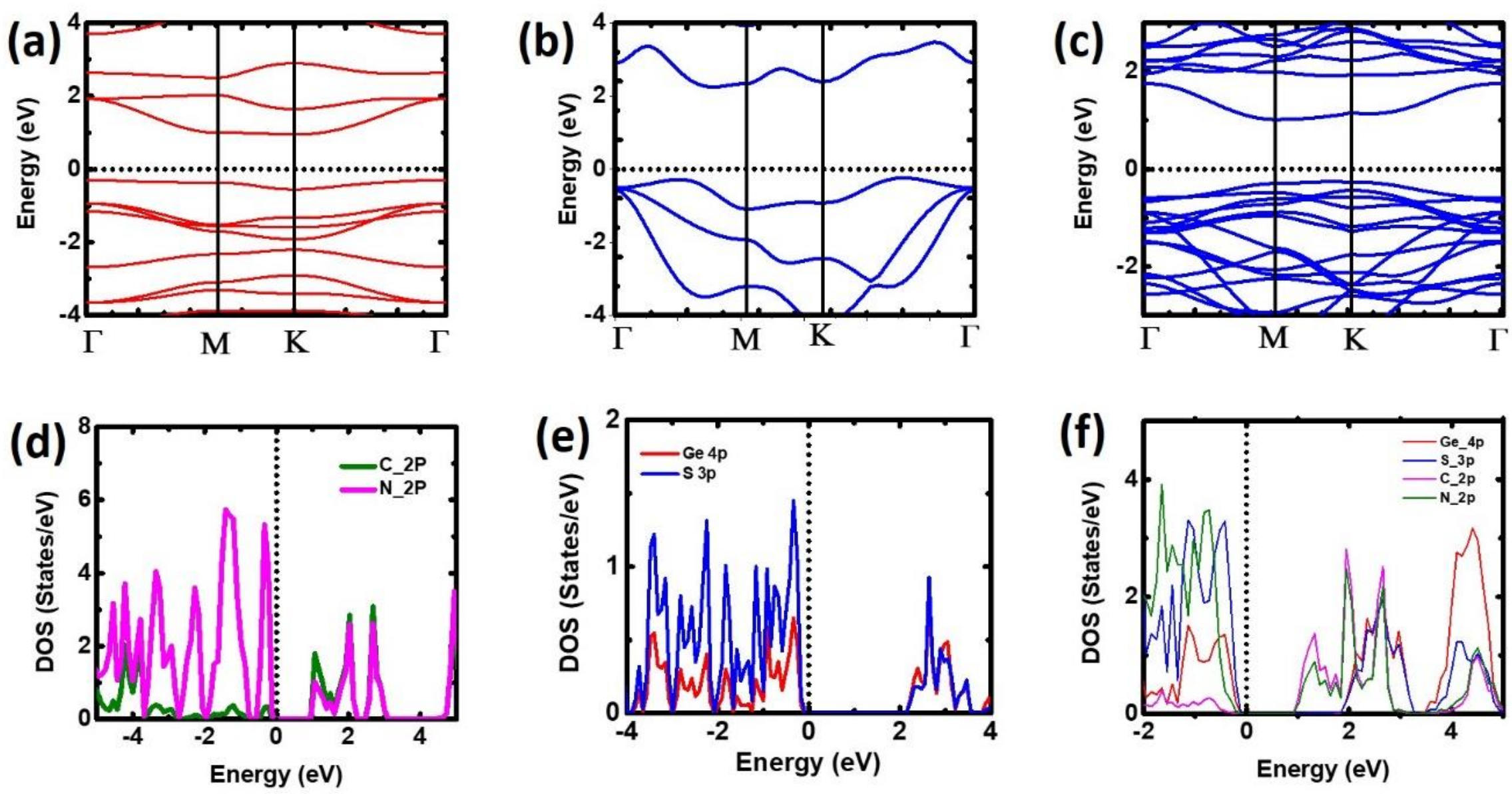

Figure S6: (**PBE functional Results**): Electronic band structure and projected density of states for (a,d) g-$C_3N_4$ monolayer, (b,e) GeS free-standing monolayers and (c,f) g-$C_3N_4$/GeS heterostructure.

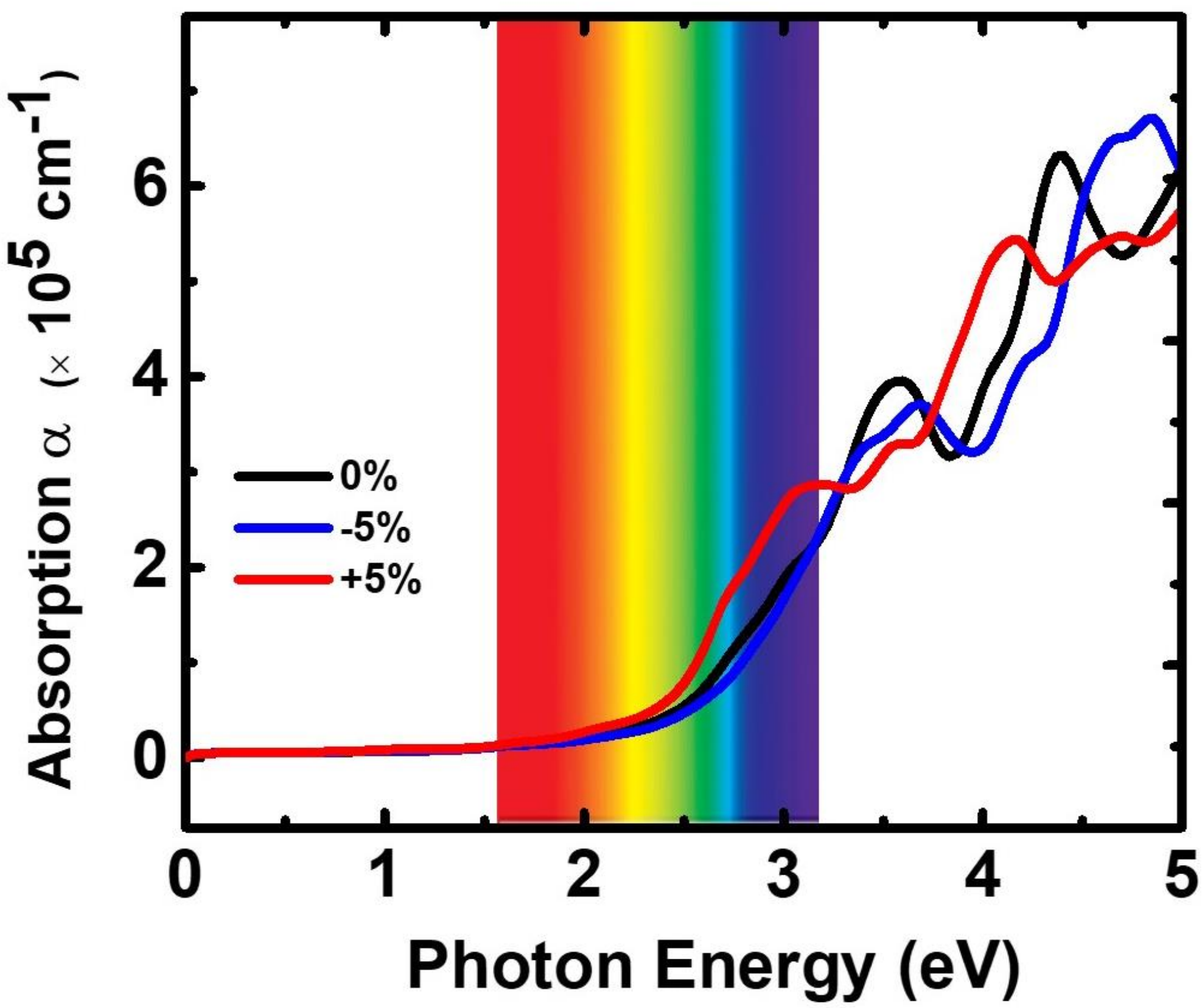

Figure S7: Optical absorption spectra of g-$C_3N_4$/GeS heterostructure as a function of biaxial strain